# Dinámica vibracional de un medio granular 3D compuesto de partículas poliédricas

# Vibrational dynamics of 3D granular media composed with polyhedral grains

Emilien Azéma[1,2], Farhang Radjaï[1], Robert Peyroux[1], Frédéric Dubois[1], Gilles Saussine[3]

(1) LMGC-CNRS – Université Montpellier 2, Montpellier, France
(2) Invesitigator invitado, Grupo de investigación en Geotecnia (GRG), Universidad de Los Andes, Bogotá, Colombia
(3) Innovation and Research Department of SNCF, Paris, France.

## Abstract

*By means of tree-dimensional contact dynamics simulations, we analyze the vibrational dynamics of a confined granular layer in response to harmonic forcing. The sample is composed of polyedric grains with a shape derived from digitalized ballast. The system involves a jammed state separating passive (loading) and active (unloading) states. We show that an approximate expression of the packing resistance force as a function of the displacement of the free retaining wall from the jamming position provides a good description of the dynamics. We study in detail the scaling of displacements and velocities with loading parameters. In particular, we find that, for a wide range of frequencies, the data collapse by scaling the displacements with the inverse square of frequency, the inverse of the force amplitude and the square of gravity. We show that the mean compaction rate increases linearly with frequency up to a characteristic frequency of 10 Hz and then it declines in inverse proportion to frequency.*

## Resumen

*En la mayoría de los estudios existentes sobre medios granulares vibrados, el material granular está compuesto de partículas esféricas o elipsoidales, y el medio es vibrado verticalmente. En este trabajo, buscamos estudiar la dinámica vibracional de un medio confinado dentro una caja con un muro sometido a un fuerza armonica horizontal, empleando el método de la Dinámica de Contactos. El medio estudiado está compuesto de partículas polihédricas cuya forma fue digitalizada a partir de agregados de balasto reales. Encontramos que el sistema pasa por tres fases : una fase bloqueada ("jamming"), una fase pasiva (descarga), y una fase activa (carga). Se muestra que es posible derivar una expresión que aproxima de forma precisa la fuerza de resistencia del medio en función del desplazamiento del muro libre. En el trabajo se estudian en detalle el desplazamiento y la velocidad del muro en función de los parámetros de carga, encontrando que, para una amplia gama de frecuencias, los desplazamientos son proporcionales al inverso de la frecuencia al cuadrado, al inverso de la amplitud de la fuerza, y a la gravedad al cuadrado. También se muestra que la velocidad de compatacción aumenta linealmente con la frecuencia, hasta una frecuencia característica de 10 Hz, y que luego disminuye de forma proporcional al inverso de la frecuencia.*

## 1 INTRODUCTION

The dynamics of dense granular materials subjected to vibrations involves collective phenomena resulting from kinematic constraints (steric exclusions, boundary and finite size effects, …) and energy dissipation (Aranson 2006). Well-known examples of the vibration-induced phenomena are compaction, convective flow, size segregation and standing wave patterns at the free surface (Knight 1995, Clement 1996, Liffman 1997, Sano 2005). Three different states can be distinguished depending on the intensity and frequency of vibrations: 1) Gas-like or fluidized state: The rate of energy input is such that there are no enduring contacts between particles and the material behaves as a dissipative gas (Jaegger 1996, Ludding 1996), 2) Solid-like state: Vibrational energy propagates through the network of enduring contacts between particles and the material undergoes slow rearrangements and progressive compaction (Kudrolli 2004). 3) Liquid-like state: Both particle migration and enduring contact networks are involved in the dynamics and various collective effects can be observed (Aoki 1996, Liffman 1997).

We may distinguish two methods for inducing vibrational dynamics: by imposed cyclic displacements of a wall or the container (shaking); or by cyclic modulation of a confining stress. The first method has been used in most experiments on granular beds (Ludding 1996, Ben-Naim 1996, Josserand 2000). In this case, the control parameters are the amplitude $a$ and the frequency $v$ of the vibrations corresponding to a maximal acceleration $a\omega^2$ where $\omega = 2\pi v$. When a material is moulded inside a closed box, the vibrations should rather be induced by varying a confining force, e.g. a force acting on a wall. Then, the amplitude of displacements is a function of the forcing frequency, and the level of particle accelerations depends on both the applied cyclic force and the reaction force of the packing (Azema 2006, 2008). Moreover, in nearly all studies, spherical or nearly spherical particles in 3D or disks or polygons in 2D have been used.

We are interested here in the evolution of the packing in the course of harmonic loading, the short-time compaction (during the first cycles) and the scaling of the dynamics with loading parameters. We used discrete-element numerical simulations by means of the contact dynamics method (CD) in 3D with rigid irregular polyhedral particles. The system is explored for a broad set of loading parameters including the frequency and amplitude of the harmonic driving force. We first introduce the numerical procedures. Then, we present in three sections the dynamics of the packing, scaling with loading parameters, and the evolution of solid fraction.

## 2 NUMERICAL PROCEDURES

The simulations were carried out by means of the contact dynamics (CD) method with irregular polyhedral particles (Jean 1992, Moreau 1994, Radjaï 1999). The CD method is based on implicit time integration of the equations of motion and a nonsmooth formulation of mutual exclusion and dry friction between particles. This method requires no elastic repulsive potential and no smoothing of the Coulomb friction law for the determination of forces. For this reason, the simulations can be performed with large time steps compared to molecular dynamics simulations. We used *LMGC90* which is a multipurpose software developed in our laboratory, capable of modeling a collection of deformable or undeformable particles of various shapes by different algorithms (Dubois 2003).

Our numerical samples are composed of rigid polyhedral particles with shapes and sizes that represent those of ballast grains (Fig. 1) (Saussine 2004). Each particle has at most 70 faces and 37 vertices and at least 12 faces and 8 vertices. A sample contains nearly 1200 particles. The particle size is characterized as the largest distance between the barycenter and the vertices of the particle, to which we will refer as "diameter" below. We used the following size distribution: 50% of diameter $d_{min} = 2.5$ cm, 34% of diameter 3.75 cm, and 16% of diameter $d_{max} = 5$ cm.

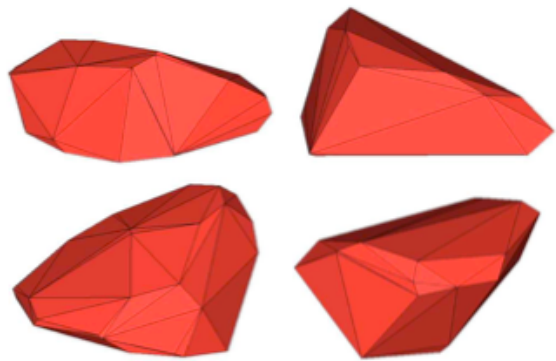

Figure 1 : Examples of polyhedral shapes used in the simulations.

The particles are placed in a rectangular box under the gravity $g$ and the upper wall is raised 1 cm and fixed ; figure 2.

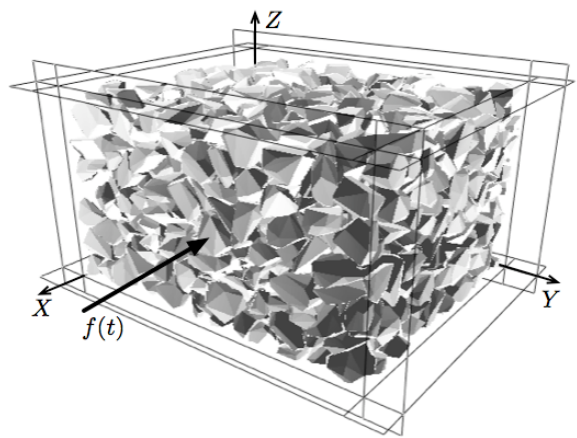

Figure 2 : A snapshot of the packing inside a box with a free wall over which the driving force $f(t)$ is applied along the $x$ direction.

The coefficient of friction between the particles and with the horizontal walls was fixed to 0.4, but it was 0 at the vertical walls. One of the walls is allowed to move horizontally (x direction in Fig. 2) and subjected to a harmonic driving force :

$$f(t) = \frac{(f_{max} + f_{min})}{2} - \frac{(f_{max} - f_{min})}{2} \sin \omega t, \quad (1)$$

where $f_{max}$ and $f_{min}$ are the largest and lowest compressive (positive) forces acting on the wall. All other walls are immobile. For all simulations the time step was $2.10^{-4}$ s.

## 3 ACTIVE AND PASSIVE DYNAMICS

If $f_{min}$ is above the (gravitational) force $f_g$ exerted by the grains on the free wall, $f$ will be large enough to prevent the wall from backward motion during the whole cycle. In other words, the granular material is in *passive state* and the major principal stress direction is horizontal. In this limit, no extension will occur following the initial contraction. On the other hand, if $f_{max}$ is below the force $f_g$ exerted by the grains, $f$ will never be large enough to prevent the extension of the packing. This corresponds to the *active state* where the major principal stress direction remains vertical. In all other cases, both contraction and extension occur during each period, and the displacement of the free wall will be controlled by $f_{min}$. Without loss of generality, we set $f_{min} = 0$. This ensures the largest possible displacement of the wall in the active state.

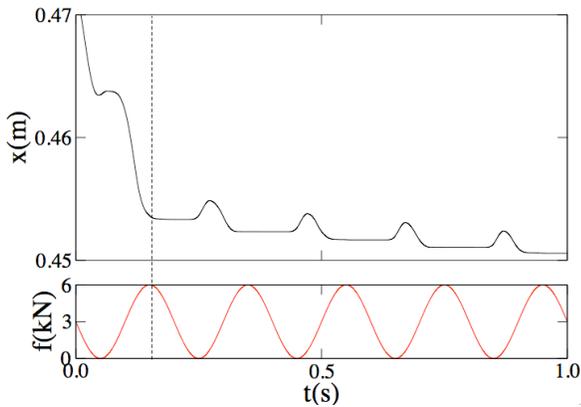

Figure 3 : The evolution of the displacement $x(t)$ of the free wall (up) in response to harmonic loading (down)

Figure 3 shows $x(t)$ for frequency $\nu = 5$ Hz over a time interval $\Delta t = 1$ s. We distinguish a fast initial contraction ($t < 1$ s) followed by slow contraction (decreasing $x$) over four periods. The initial contraction is a consequence of the gap left between the free surface of the packing and the upper wall. This initial volume change is almost independent of frequency. The motion of the free wall is governed by the equation of dynamics,

$$f - f_g = m_w \ddot{x}, \qquad (2)$$

where $m_w$ is the mass of the wall. To predict the motion of the free wall, we need to express the force $f_g$ as a function of $x$.

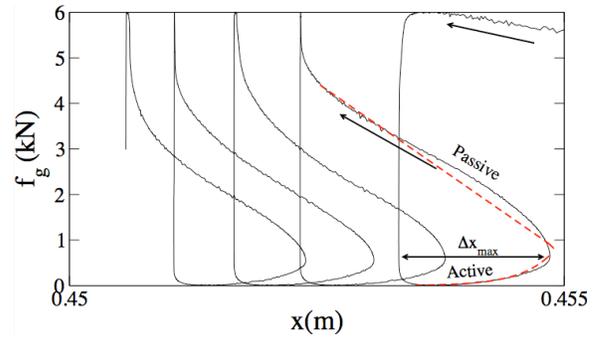

Figure 4 : Force $f_g$ exerted by the particles on the free wall as a function of displacement $x$ (full line) and an approximate fitting form (dashed line), see text.

Figure 4 shows the horizontal force $f_g$ exerted by the packing on the wall as a function of $x$ over four periods. In the active phase, $f_g$ grows slightly exponentially with $x$ (dashed line in Fig. 4). The maximum displacement $\Delta x_{max}$ occurs at the end of this phase. In the passive phase, $f_g$ grows faster and almost linearly as $x$ decreases. The vertical line corresponds to the jammed state where $f_g$ decreases with $f$, in this case we have $f_g = f$.

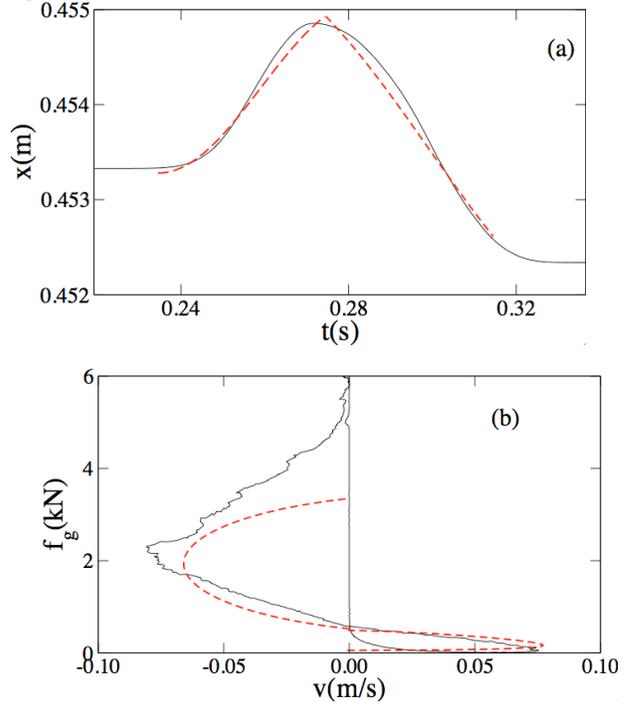

Figure 5 : Displacement $x$ of the free wall as a function of time (a) and $f_g$ vs $v$ (b) over one period (full line) and analytical fit from the phenomenological model (dashed line).

Figure 5a and 5b shows the evolution of the position $x$ as a function of time and the variation of $f_g$ as a function of $v$ for one period, together with the solution of the model. Excluding jamming and unjamming transients, the analytical solution provides a fairly good approximation for the simulation data although the largest contraction velocity is under-estimated in the passive state.

Although we focus here on the average dynamics of the packing, i.e. the displacements of the free wall, it is important to note that the grain velocity field is not a simple oscillation around an average position. The grains undergo a clockwise convective motion in the cell as shown in Figure 5.

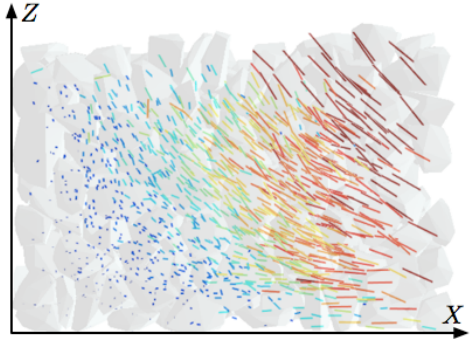

Figure 6 : Instantaneous particle velocity field in the passive state, i.e. during inward motion of the free wall.

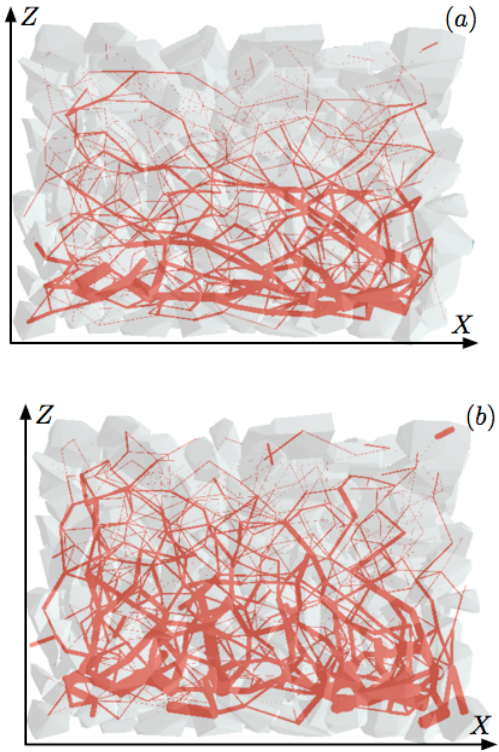

Figure 7 : Normal forces in the passive (a) and active (b) states in a section of the packing. The segments connect particle centers with a thickness proportional to the normal force.

On the other hand, the contact forces evolve between a fully jammed state, where nearly horizontal force chains dominate ; Figure 7a, and the active state, where nearly vertical gravity-induced chains can be observed ; Figure 7b.

## 4 SCALING WITH LOADING PARAMETERS

We performed a series of simulations for frequencies $\nu$ ranging from 1 Hz to 60 Hz and for a total time of 1 s. All simulations yield similar results for dynamics. Moreover, a simple dimensional analysis leads to the collapse of the data on a single plot. Indeed, the frequency sets the time scale $\tau = \nu^{-1}$. Force scales are set by the largest driving force $f_{max}$ in the passive state and the particle weights $mg$ as well as the smallest driving force $f_{min}$ in the active state. Hence, dimensionally, for fixed values of $mg$, $f_{min}$ and $f_{max}$, all displacements are expected to scale with $\nu^{-2}$ and all velocities with $\nu^{-1}$.

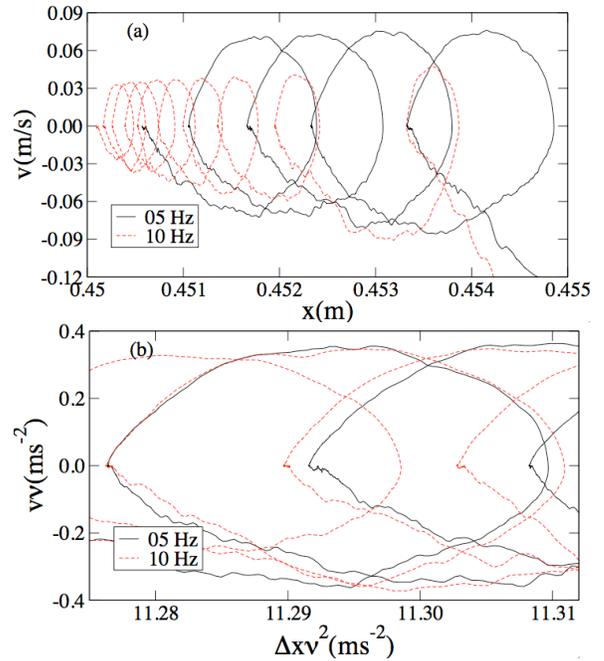

Figure 8 : Phase space trajectories for two frequencies without scaling (a) and with scaling (b) of the displacements and velocities with respect to the frequency.

This scaling is shown in Figure 8 where the phase space trajectory is shown for $\nu = 5$ Hz and $\nu = 10$ Hz without scaling and after scaling the displacements $\Delta x$ by $\nu^{-2}$ and the velocities $v$ by $\nu^{-1}$. We see that the data from both simulations collapse nicely on the same trajectory after scaling. To check directly this scaling, in Figure 9 we have plotted the maximum displacement $\Delta x_{max}$ in the active state and the maximum velocity $v_{max}$ in the passive state as a function of $\nu$. The corresponding fits are excellent.

The role of force parameters $mg$, $f_{min}$ and $f_{max}$ is less evident. Since we have $f_{min} = 0$, we expect $\Delta x_{max}$ to be dependent on the ratio $mg/f_{max}$ representing the relative importance of the gravitational to driving forces. Indeed, our data show that $\Delta x_{max}$ varies as $f_{max}^{-1}$; Figure 10.

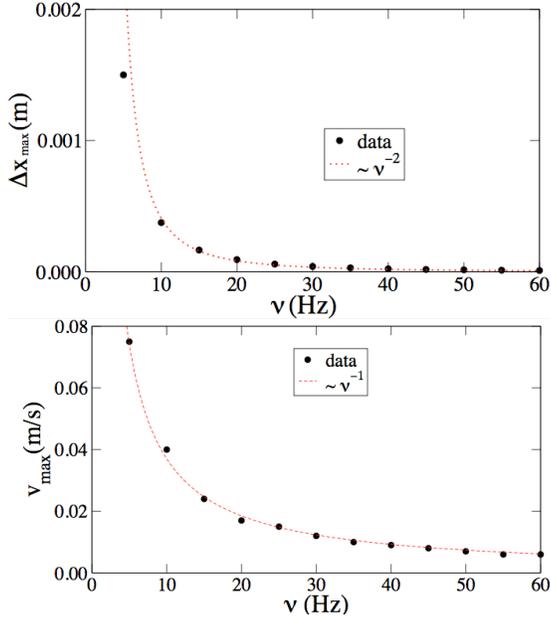

Figure 9 : Maximum displacement $\Delta x_{max}$ (up) and the maximum velocity $v_{max}$ (down) as a function of the frequency $\nu$

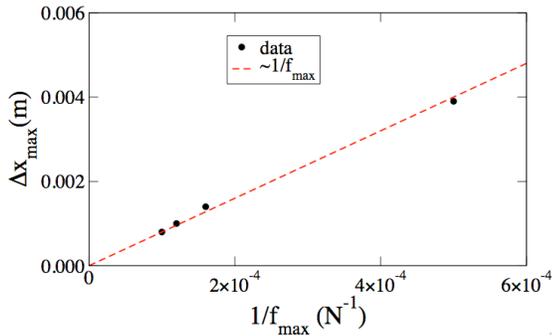

Figure 10 : Scaling of the maximum displacement $\Delta x_{max}$ with the force amplitude $f_{max}$.

Hence, we propose the following expression for the scaling of displacements with loading parameters:

$$\Delta x_{max} = C \left(\frac{m}{m+m_w}\right) \left(\frac{mg}{f_{max}}\right) \left(\frac{g}{\nu^2}\right) \quad (3)$$

where $C$ is a dimensionless prefactor. Figure 11 shows $\Delta x_{max}$ as a function of (3) from different simulations with different values of $\nu$, $f_{max}$, $g$ and $m_w$. We see that the data are in excellent agreement with Eq. 3. The prefactor is $C \sim 0.01$. This scaling is the same as in 2D simulations with a material constant $C \sim 0.05$ for polygonal particles (Azéma 2006). Let us also remark that Eq. 3 predicts that $\Delta x_{max}$ varies as $g^2$. This prediction agrees well with our simulation data

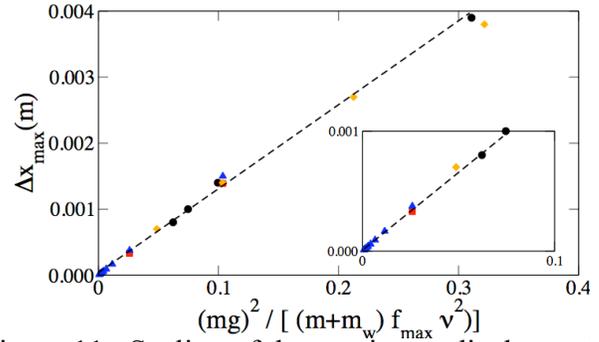

Figure 11 : Scaling of the maximum displacement $\Delta x_{max}$ with loading parameters from simulations with different values of the frequency (triangles), the force amplitude (circles), the gravity (squares), and for the mass (diamonds) of the free wall. The inset shows the plot near the origin

## 5 COMPACTION AND COMPACTION RATE

In order to evaluate the solid fraction $\rho$, we consider a control volume enclosing a portion of the packing inside the simulation cell. This volume does not include the initial gap between the top of the packing and the upper wall. The initial value of the solid fraction is 0.50 and, since the grains are angular-shaped, its variations $\Delta\rho$ from the initial state are large.

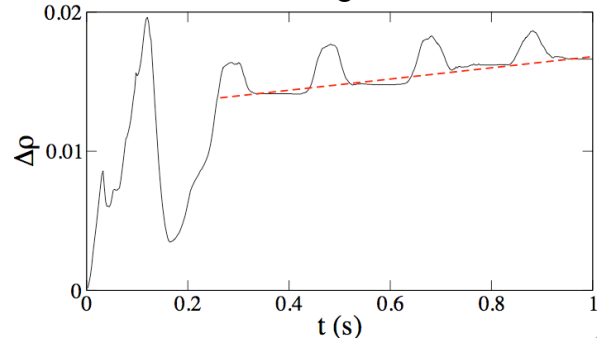

Figure 12 : Evolution of the solid fraction $\Delta\rho$ from the initial state as a function of time over several periods for 5 Hz.

Figure 12 shows the evolution of the variation $\Delta\rho$ of solid fraction for several periods. An initial compaction of 2% is followed by oscillations with a small increase of $\Delta\rho$ in each period. The initial compaction should be attributed to the initial state where the packing is not yet fully confined. We use $\rho_0 = 0.52$, reached after a time lag of 0.2 s, as the reference value for the evolution of solid fraction. The compaction of the packing slows down logarithmically at long times (Deboeuf 2005). But, the short-time compaction can well be approximated by a linear function with a constant compaction per period $\Delta\rho_1$ as seen in figure 12. The average compaction rate $\dot{\eta}$ over several periods and for a total time interval $\Delta t$ is then:

$$\dot{\eta} = \frac{1}{\rho_0}\frac{\Delta\rho}{\Delta t} \sim \frac{\Delta\rho_1}{\rho_0}\nu \qquad (4)$$

Figure 13 shows $\dot{\eta}$ as a function of $\nu$. We see that only at low frequencies, $\dot{\eta}$ increases linearly with $\nu$. Beyond a characteristic frequency $\nu_c$, $\dot{\eta}$ declines with $\nu$. The largest compaction rate $\dot{\eta}_{max}$ occurs for $\nu_c \sim 10$ Hz. The corresponding time $\tau_c = \nu^{-1}$ represents the characteristic time for the relaxation of the packing. In the active state, the packing needs a finite rearrangement time to achieve a higher level of solid fraction. As long as the vibration period $\tau = \nu^{-1}$ is longer than $\tau_c$, the packing has enough time to relax fully to a more compact equilibrium state. But, when the period $\tau$ is below $\tau_c$, the relaxation is incomplete.

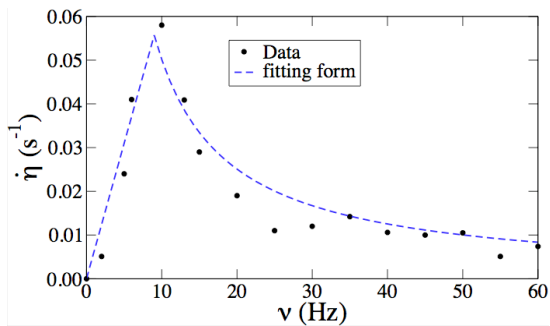

Figure 13 : The compaction rate $\dot{\eta}$ as a function of the frequency (circles) with an approximate fitting form.

It is expected that $\Delta\rho_1$ should follow the same scaling with the frequency as the displacement of the retaining wall, i.e. $\Delta\rho_1 \sim \Delta\rho_{max}\nu^{-2}$, where $\Delta\rho_{max}$ is the largest compaction between two periods. Hence, from Equation 4 and imposing the continuity at $\nu = \nu_c$, we get :

$$\begin{cases} \frac{\Delta\rho_{max}}{\rho_0}\nu & \nu < \nu_c, \\ \frac{\Delta\rho_{max}}{\rho_0}\nu_c^2\,\nu^{-1} & \nu > \nu_c. \end{cases} \qquad (5)$$

This form is plotted in Figure 13 together with the data points. It is remarkable that, although $\nu_c$ is the only fitting parameter, the compaction rate $\dot{\eta}$ is well adjusted by Equation 5. The prefactor $\Delta\rho_{max}/\rho_0 \sim 0.0025$.

The characteristic time $\tau_c \sim 0.1$s is of the same order of magnitude as the time required for one particle to fall down a distance equal to its diameter. Obviously, the above findings concern only short-time compaction. At longer times, $\dot{\eta}$ declines with time, but the scaling with frequency according to Equation 5 is expected to hold at each instant of evolution of the packing.

## 6 CONCLUSION

In this paper, we analyzed the short-time behavior of a constrained granular system subjected to vibrational dynamics. The vibrations are induced by harmonic variation of the force exerted on a free retaining wall between zero and a maximum force. The system as a whole has a single degree of freedom represented by the horizontal position of the free wall. This system involves a jammed state separating passive (loading) and active (unloading) states. The contact dynamics method was employed to simulate and analyze the dynamics of this system composed with polyhedral particles.

By construction, our system is devoid of elastic elements and, hence, the behavior is fully governed by collective grain rearrangements.

In the loading phase, the reaction force (exerted by the grains on the free wall) rises almost linearly with the displacement of the free wall, but it increases considerably at the end of this phase in transition to the jammed state. This force enhancement features the jamming transition compared to the rest of the passive state. The reaction force decreases then in the jammed state, balancing thus exactly the driving force, until the latter is low enough for the grains to push the free wall away under the action of their own weights. This unjamming process occurs smoothly and the reaction force increases only slightly but exponentially during the unloading phase. We showed that a rough expression of the reaction force as a function of the displacement of the free wall with respect to the jamming position provides a good prediction of the dynamics except at the jamming and unjamming transients.

Dimensional analysis was used to scale the displacements with the frequency of oscillations. It was shown that the data collapse by scaling the displacements by the inverse square of frequency. We also studied the scaling with confining force and particle weights.

We also investigated the compaction rate of our numerical samples. It is nearly constant for short times. It was shown that the compaction rate increases linearly with frequency up to a characteristic frequency and then it declines nearly in inverse proportion to frequency. The characteristic frequency was interpreted in terms of the time required for the relaxation of a packing in each period to a more compact state by collective grain rearrangements under the action of gravity.

REFERENCES

Aranson, I.S, Tsimring. (2006). "*Patterns and Collective Behavior in Granular Media, Theoretical Concepts*" Reviews of Modern Physics, 78, 641-692.

Knight, J.B., Fandrich, C.G., Lau C.N., Jaeger H. Nagel S. (1995). "*Density relaxation in a vibrated granular material*" Phys. Rev. E, 51, 3957.

Clement E., Vanel L., Rajchenbach J., Duran J. (1996). "*Pattern formation in vibrated granular layer*" Phys. Rev. E, 53, 2972.

Liffman K., Metcalfe G. Cleary P. (1997). "*Granular convection and transport due to horizontal shaking*" Phys. Rev. Lett., 79, 4574-4576.

Sano O. (2005). "*Dilatancy, buckling, and undulations on a vertically vibrating granular layer*" Phys. Rev. E., 5, 051307.

Jaegger H.M., Nagel S.R. Behringer R.P. (1996). "*Granular solid, liquids and gases*" Reviews of Modern Physics, 58, 1259-1273.

Ludding S., Nagel S.R. Behringer R.P. (1996). "*Granular materials under vibrations : simulation of rotating spheres*" Phys. Rev. E, 52, 4442 - 4457.

Kudrolli A. (2004). "*Size separation in vibrated granular matter*" Rep. Prog. Phys., 67, 209 - 247.

Aoki K.M. Akiyama T. (1996). "*Spontaneous Wave Patterns Formation in Vibrated Granular Materials*" Phys. Rev. Lett., 20, 4166-4169.

Ben-Naim E., Knight J.B., Nowak E.R., (1996). "*Slow relaxation in granular compaction*" J. Chem. Phys., 100, 6778.

Josserand C., Tkachenko A.V., Mueth D.M., Jaeger H.M.,(2000). "*Memory Effects in Granular Materials*" Phys. Rev. E, 85, 3632 - 3635.

Azéma E., Radjaï F., Peyroux R., Dubois F. Saussine G., (2006). "*Vibrational dynamics of confined granular material*" Phys. Rev. E, 74, 031302.

Azéma E., Radjaï F., Peyroux Richefeu V., G. Saussine G., (2008). "*Short-time dynamics of a packing of polyhedral grains under horizontal vibrations*" Eur. Phys. J. E, 26, 327-335.

Moreau J.J. (1994). "Some numerical methods in multi dynamics : application to granular materials", Eur. J. Mech. A Solids, 93-114.

Jean M., Moreau J.J, (1992). "*Unilaterality and dry friction in the dynamics of rigid body collections in Proceedings of Contact Mechanics International Symposium*", Presses Polytechniques et Universitaires Romandes (Lausanne, Switzerland), 31-48.

Radjaï F. (1999). "Multicontact dynamics of granular systems", Computer physics communications, 121-122, 294.

Dubois F., Jean M. (2003). " LMGC90 une plateforme de développement dédiée à la modélisation des problèmes d'intéraction", Actes du sixième colloque national en calcul des structures, volume1, CSMA-AFM-LMS, 2003

Saussine G. (2004) *"Contribution à la modélisation de granulats tridimensionnels: application au ballast, Phd dissertation,"* Université Montpellier 2, France, 2004. http://tel.archives-ouvertes.fr/tel 00077519/en/

Deboeuf S., Dauchot O., Staron L., Mangeney A., Vilotte J.P (2006). "*Memory of the unjamming transition during cyclic tiltings of a granular pile*" Phys. Rev. E, 72, 051305.